%% file: main.tex
\documentclass[10pt,a4paper,conference]{IEEEtran}
\ifCLASSINFOpdf
  \usepackage[pdftex]{graphicx}
  \usepackage[bookmarks=false]{hyperref}
\else
  \usepackage[dvips]{graphicx}
\fi
\ifCLASSOPTIONcompsoc
  \usepackage[caption=false,font=normalsize,labelfont=sf,textfont=sf]{subfig}
\else
  \usepackage[caption=false,font=footnotesize]{subfig}
\fi

\usepackage{comment} 
\usepackage{amsmath}

\usepackage[T1]{fontenc}

\usepackage[usenames,dvipsnames,svgnames]{xcolor}

\newif\ifcommentson

\newcommand{\optional}[1]{\ignorespaces}

\begin{document}
%
\title{Implementation of Virtual Network Function Chaining through Segment Routing\\in a Linux-based NFV Infrastructure
}

\author{\IEEEauthorblockN{
Ahmed AbdelSalam\IEEEauthorrefmark{1},
Francois Clad\IEEEauthorrefmark{4},
Clarence Filsfils\IEEEauthorrefmark{4},
Stefano Salsano\IEEEauthorrefmark{2},
Giuseppe Siracusano\IEEEauthorrefmark{2},
Luca Veltri\IEEEauthorrefmark{3},
}
\IEEEauthorblockA{
\IEEEauthorrefmark{1}Gran Sasso Science Institute,
\IEEEauthorrefmark{2}University of Rome Tor Vergata,
\IEEEauthorrefmark{3}University of Parma,
\IEEEauthorrefmark{4}Cisco Systems
}
\\
\textbf{Extended version of the conference paper~\cite{vnf-chain-srv6}  - v04 - April 2017}
}


%


\maketitle

\input{inc/abstract}

%
\IEEEpeerreviewmaketitle

\vspace{2ex}


\input{inc/introduction}

\input{inc/architecture}

\input{inc/implementation}

\input{inc/results}

\input{inc/soa}

\input{inc/conclusions}





%

\bibliographystyle{IEEEtran}
\bibliography{main}

\end{document}

%% file: inc/abstract.tex
\begin{abstract}

In this paper, we first introduce the NFV architecture and the use of IPv6 Segment Routing (SRv6) network programming model to support Service Function Chaining in a NFV scenario. We describe the concepts of SR-aware and SR-unaware Virtual Network Functions (VNFs). The detailed design of a network domain supporting VNF chaining through the SRv6 network programming model is provided. The operations to support SR-aware and SR-unaware VNFs are described at an architectural level and in particular we propose a solution for SR-unaware VNFs hosted in a NFV node. The proposed solution has been implemented for a Linux based NFV host and the software is available as Open Source. Finally, a methodology for performance analysis of the implementation of the proposed mechanisms is illustrated and preliminary performance results are given.

\end{abstract}

%% file: inc/introduction.tex
\section{Introduction}
\label{sec:introduction}

\let\svthefootnote\thefootnote
\let\thefootnote\relax\footnotetext{This work was partly performed in the context of the Superfluidity project, which received funding from the EU Horizon 2020 programme under grant agreement No. 671566}
\let\thefootnote\svthefootnote


The concept of Network Function Virtualization (NFV)~\cite{mijumbi2015network} is reshaping the way in which telecommunication networks and services are designed and operated. Traditional network functions are transformed in \textit{VNFs} (Virtual Network Functions), running over a distributed, cloud-like infrastructure referred to as \textit{NFVI} (NFV Infrastructure). In the NFV approach, services are implemented by properly chaining VNFs that can be distributed over the NFVI. This process is called \textit{Service Function Chaining} (SFC). An overview of the issues related to the deployment and chaining of VNFs is reported in~\cite{RFC7498}, while the SFC architecture as standardized by the IETF is included in~\cite{RFC7665}. Citing from~\cite{RFC7498}, VNFs can act at various layers of a protocol stack (e.g., at the network layer or other layers) and examples of service functions include: firewalls, WAN and application acceleration, Deep Packet Inspection (DPI), server load balancers, NAT44, NAT64, HTTP header enrichment functions, TCP optimizers. In general, it is possible to chain both virtualized functions (VNFs) and physical nodes in a Service Function Chain. Hence, in the text above we should have referred to \textit{service functions} rather than to VNFs. For the sake of simplicity, and considering the trend towards network softwarization, we only refer to VNFs throughout the paper.

The SFC architecture defined in~\cite{RFC7665} is a high level one. The generic concept of \textit{SFC encapsulation} is introduced, without specifying the protocol mechanisms that are needed to enforce the forwarding of packets along the chain of VNFs. The definition of a new header called Network Service Header (NSH) that can be inserted into packets or frames to support SFC is proposed in~\cite{nsh-id}. On the other hand, in this paper we follow the approach of using the \textit{Segment Routing} (SR)~\cite{filsfils2015segment} architecture to support SFC (see~\cite{lebrun2015leveraging}), considering in particular, the IPv6-based Segment Routing (SRv6)~\cite{lebrun2016design}. The SR architecture relies on the \textit{source routing} paradigm. A node can add to a packet an ordered list of instructions, denoted as \textit{segments}, that can be used to steer the packet through a set of intermediate steps in the path towards its final destination. Although the SR architecture can operate over a MPLS or an IPv6 data plane, here we only consider the IPv6 solution. In the IPv6 case, the list of segments is transported in a new type of Routing Extension Header called SR Header (SRH)~\cite{ID-ipv6-SRH}. In~\cite{srv6-net-prog}, the IPv6 Segment Routing concept is extended from the simple steering of packets across nodes to a general network programming approach. In fact, thanks to the huge IPv6 addressing space, it is possible to encode \textit{instructions} and not only \textit{locations} in a segment list. The architectural approach and the implementation described in this paper are based on the network programming model proposed in~\cite{srv6-net-prog}.  

The VNFs can be divided into two classes with respect to their interaction with the SR: \textit{SR-aware} functions and \textit{SR-unaware} functions. SR-aware functions can process the information contained in the SRH of incoming packets and can use the SRH to influence the processing/forwarding of the outgoing packets. In particular, SR-aware VNFs could directly process the SRH in IP packets or they could interact with the Operating System or with SR modules in order to read and/or set the information contained in the SRH. SR-unaware VNFs are not capable to understand the SRH, they can only reason in terms of traditional IP operations. The typical case for SR-unaware VNFs is the case in which a pre-existing VNF (also referred to as a \textit{legacy} VNF) is used in a SR-based SFC scenario. In this case, legacy VNFs need to be inserted in the SFC processing chain in such a way that they can receive, process and forward plain IP packets with no knowledge of the SRH and of the SFC infrastructure. In this case, the SFC infrastructure needs to take care of handing the packets to the SR-unaware VNFs and to receive the packets from them, performing the adaptation with the SR-based SFC processing chain.

In this paper, we consider a SRv6-based SFC scenario in a NFV infrastructure, as described in Section~\ref{sec:architecture}. We focus on the issue of steering traffic within a Linux-based NFV host that supports a potentially large number of VNFs. Our first contribution is to describe a solution for a particular class of SR-unaware VNFs. The solution is able to support a set of VNFs running as containers on the Linux NFV host. As a second important contribution, in Section~\ref{sec:implementation} we present an open-source implementation of the proposed solution, working as a Linux kernel module. The implemented module is able to adapt the behavior of SR-unaware VNFs to the needs of the SRv6-based SFC processing chain. Our testbed and a methodology for performance analysis represent the third contribution of the paper and are reported in Section~\ref{sec:methodology}. Finally we show a preliminary performance evaluation of our solution in  Section~\ref{sec:results}.

%% file: inc/architecture.tex
\section{NFV/SR Architecture}
\label{sec:architecture}
 
\begin{figure}[htpb]
	\centering
	\includegraphics[width=0.48\textwidth]{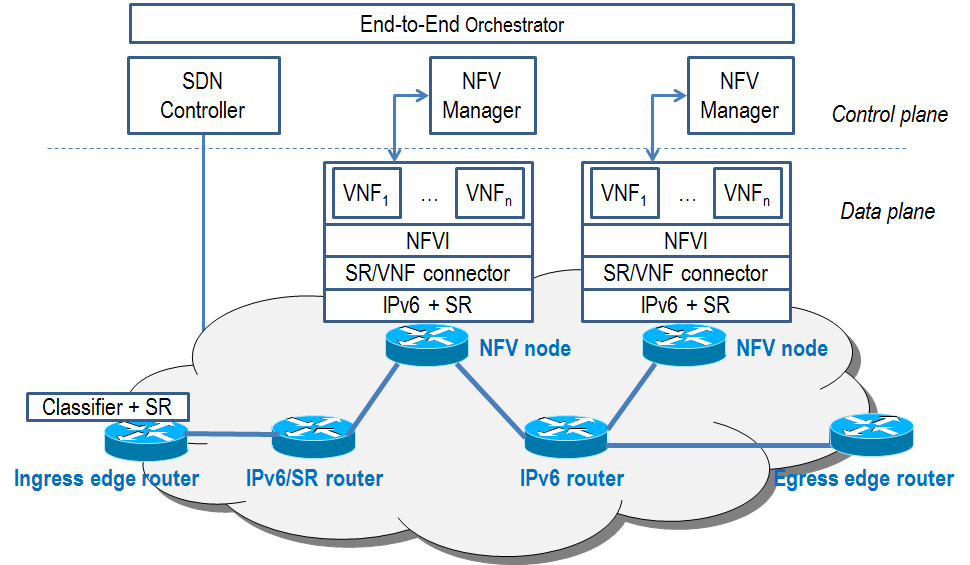}
	\caption{NFV/SR Architecture.}
	\label{fig:net-arch}
\end{figure}

In this section we describe the architecture of a network domain supporting VNF chaining through IPv6-based SR. The reference scenario is depicted in Fig.~\ref{fig:net-arch}. In the data plane, the network is internally composed of IPv6 core routers (CRs), some of which are SR-enabled, that is they are able to process the IPv6 SRH. There could be also legacy IPv6 CRs that simply forward packets regardless of the presence of the SRH (according to the current IETF draft on IPv6 SR architecture~\cite{ID-ipv6-spring}). On some nodes, referred to as NFV nodes, it is possible to run the VNFs. The set of nodes on which the VNFs can be instantiated is referred to as NFVI (Network Function Virtualization Infrastructure). In Fig.~\ref{fig:net-arch}, the NFV nodes are IPv6/SR routers that are also capable of running VNFs. Another possibility (not shown in the figure) is to have a NFV node external from the IPv6/SR router, i.e. running as a host attached to router.

We assume that a VNF instance running in a NFV node is uniquely identified by an IPv6 address. A NFV node will be able to host a number of VNF instances, and the IPv6 routing in the network can be easily configured in order to forward toward the NFV node all packets destined to the VNF instances running in the NFV node. 

According to the SFC architecture document (\cite{RFC7665}), a Service Function Chain is an ordered set of \textit{abstract} service functions that should be applied to a packet (or flow). The concrete list of Service Function instances to be traversed (including their addressing information) is referred to as \textit{Rendered Service Path} (RSP). In this paper we will use the notation \textit{VNF chain} to identify the ordered set of VNF instances to be traversed (i.e. corresponding to the RSP defined in~\cite{RFC7665}). Therefore, we will represent a VNF chain as $<$v-1,v-2,...,v-n$>$, where v-i is the IPv6 address of the i-th VNF in the chain.  

As we exploit the IPv6 Segment Routing solution, each VNF IPv6 address corresponds to a Segment IDentifier (SID) and the VNF chain can be represented in a SRH (Segment Routing Header) containing a SR path (i.e. the ordered list of segments).

Coming back to Fig.~\ref{fig:net-arch}, at the border of the domain there are the edge routers (ERs) that classify incoming packets associating them to the VNF chains (represented as lists of IPv6 addresses to be inserted in the SRH). As described in~\cite{ID-ipv6-SRH}, this is done by encapsulating the original IPv6 packets in an outer IPv6 packet with the SRH routing extension. In other words, the original packet is inserted as payload of a new packet composed of a new IPv6 header, the SRH extension that represents the VNF chain, and the original packet. The new IPv6 header has the ingress ER as IPv6 source address, the next VNF in the chain as IPv6 destination address, and the egress ER as last segment in the segment list.

When a SRH-provided packet arrives to a NFV node, with the destination address equals to one of the VNF addresses associated to the node, the packet is processed by a SR/VNF connector (see later) and then passed to the corresponding VNF. After the packet has been successfully processed by the VNF, it is passed back to the SR/VNF connector and to the underlying networking layer for being forwarded to the next VNF, to the next-hop SR router, or to the final destination.

In the control plane, the architecture includes the End-to-End Orchestrator that interacts with the NFV Managers for configuring and administrating the VNFs, and with a SDN Controller for configuring network nodes. In particular, two main tasks can be performed by the SDN controller.
\begin{enumerate}
    \item Configuration of the classifier and SR module of the ingress ERs (new classification rules and proper SR paths specifying the requested chain of VNFs), each time a new traffic flow is added to the network. Thanks to the SR approach, the per-flow configuration state is only stored in the ingress node (ER), and no per-flow modification of the rest of the network is needed.
    \item Configuration of the routing tables of the nodes according to the topology. If there are no specific requirements for traffic engineering and/or routing constraints, this function can be replaced by standard routing protocols, like OSPF. This function is not executed on a per-flow basis, since it is requested only when a new node is added or removed, or when some modification of the routing is needed. 
\end{enumerate}

As shown in Fig.~\ref{fig:net-arch}, inside each NFV node the SR/VNF connector is the module in charge of logically connecting the SR routing with local VNFs. This operation can be logically split into three phases: (i) identification of the target VNF and (potentially) modifications of the packet, in order to let the packet be correctly processed by the legacy VNFs (see below); (ii) dispatching the packet to the proper VNF; iii) restoring the correct SR encapsulation after the VNF returned the packet, in case modifications had been applied previously to the packet by the SR/VNF connector.

In a simple scenario, VNFs are instantiated in the NFV nodes by the Orchestrator in a semi-permanent way, based on static configuration. In a more advanced scenario, VNFs could instantiated ``on-the-fly'' when the first packet, belonging to a new flow and targeted to a given VNF, arrives at the node; the SR/VNF connector is also in charge of the instantiation operation. In this paper, we only consider the basic scenario with static instantiation of VNFs.

Regarding the VNFs and the operation performed by the SR/VNF connector, two scenarios are possible as described in the next subsections: (a) the VNFs are SR-aware, or (b) the VNFs are SR-unaware. According to the SRv6 network programming model~\cite{srv6-net-prog}, the operations to be performed are associated with the Segment IDentifier in the SR path. In other words, the IPv6 address used to include a VNF in a service chain does not only identify the VNF instance, but also instruct the NFV node to perform the operations (like decapsulation) that are needed before handing the packet to the VNF.

\subsection{SR-aware network functions}

In this case, VNFs are aware of SR and the corresponding IPv6 SR packet encapsulation. This means that the VNFs are able to process the original packet despite the fact that it has been modified with the SR encapsulation. Since these types of VNFs are aware of SR and SRH, they may be also able to manage the SRH by: (i) reading the sequence of identifiers of the past VNFs that have already processed the packet and the sequence of identifiers of the following VNFs that still have to process the packet, (ii) changing the chain of the following VNFs by adding, removing, and possibly reordering the list. This opens the possibility of advanced VNF operations.

When a node runs SR-aware VNFs, it acts as follows:
\begin{itemize}
    \item when a SR-encapsulated packet arrives to the node, the destination address and the SRH are processed by the IPv6 layer;
    \item if the destination address corresponds to the current node (or to any VNFs running within the node), the destination address and the SRH are updated according to the IPv6 SRH protocol;
    \item if the address corresponds to a local VNF the packet is passed to the SR/VNF connector;
    \item the SR/VNF connector passes the packet to the proper VNF; 
    \item the packet is then processed by the VNF; the packet can be processed: (i) without any change in the packet header and payload, (ii) with modification, or (iii) it can be dropped;
    \item in the first two cases (packet successfully processed, i.e., not dropped) more VNFs can be enforced by the VNF for further processing by modifying the segment list in the SRH (see later for more details);
    \item the packet is returned to the SR/VNF connector, and to the IPv6 layer; the IPv6 layer processes the packet and if the destination address corresponds to a VNF of this node the above steps are executed again; otherwise, the packet is forwarded to the next segment. In case the current node is the last segment the SR encapsulation is removed and the original packet is processed and forwarded.
\end{itemize}

When different VNFs are running in the same NFV node and are part of the VNF chain associated to a given packet, the previous steps are executed more than once in the same NFV node for the packet. In this case, the SR/VNF connector could directly resend the packet towards the next VNF, rather than sending back the packet to the IPv6 layer as described above.  

When a VNF is aware of the SR encapsulation, it could be able to modify the segment list in the SRH (adding, removing, or reordering the next VNFs in the list). Three cases are possible: 1) the VNF is only allowed to insert new VNFs between the current VNF and the next one (next segment); 2) new VNFs can be inserted in any position along the VNF chain (SR path); 3) the current VNF is allowed to fully modify the segment list by adding, removing, and/or re-ordering next VNFs (segments). The last two cases require that the VNF is aware of (i.e. it knows) the VNFs corresponding to the segment IDs already present in the segment list. In all these three cases, the SR/VNF connector is in charge of controlling that the segment list in the SRH has been modified correctly, according to the access and security rules given to the VNFs.

\subsection{SR-unaware network functions}

A VNF is SR-unaware if it is not able to process an incoming packet enveloped by the SR encapsulation, and hence it is not able to recognize the original packet from the SR-encapsulated packet.
In this case, in order to correctly apply the VNF to the original packet, the SR/VNF connector must pre-process the packet by removing the SR encapsulation, and re-apply it when the packet is returned by the VNF.

A node with SR-unaware VNFs acts as follows:
\begin{itemize}
    \item when a SR-encapsulated packet arrives to the node, the destination address and the SRH are processed by the IPv6 layer;
    \item if the destination address corresponds to the current node (or to any VNFs running within the node), the destination address and the SRH are updated according to the IPv6 SRH protocol;
    \item if the address corresponds to a local VNF, the packet is passed to the SR/VNF connector; the SR/VNF connector processes the SR-encapsulated packet and extracts the original packet; the detached outer IPv6 header including the SRH is, in some way, stored in order to be re-attached to the packet when it is returned by the NF;
    \item the SR/VNF connector passes the new packet to the proper VNF;
    \item the packet is processed by the VNF; the packet can be accepted or dropped; if accepted, the packet is captured back by the SR/VNF connector; the packet returned by the VNF can be the same that as been sent to the VNF or a new one (modified by the VNF);
    \item the SR/VNF connector retrieves the original outer IPv6 header including the SRH of the packet, and re-attaches it to the packet;
    \item the packet is then returned to the IPv6 layer and processes; if the destination address corresponds to a VNF of this node, the above steps are executed again, otherwise the packet is forwarded.
\end{itemize}

\begin{figure}[htpb]
	\centering
	\includegraphics[width=0.48\textwidth]{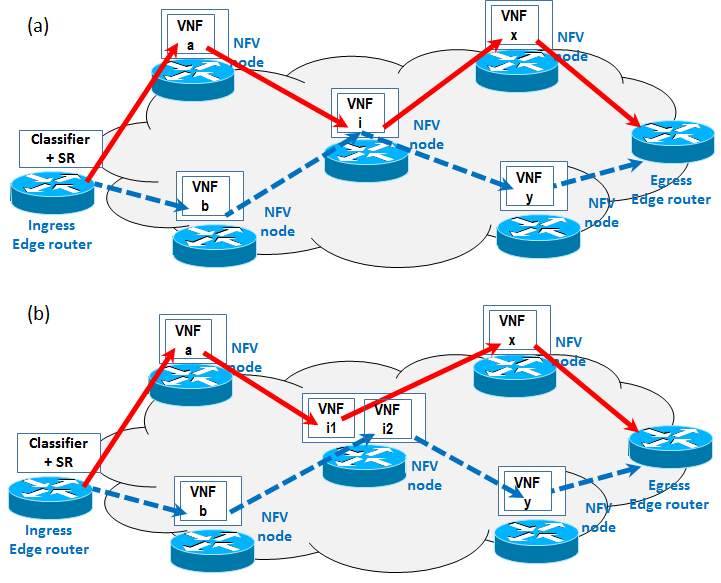}
	\caption{Service Function Chains examples}
	\label{fig:two-chains}
\end{figure}

According to the above described operations, when the SR/VNF connector receives a packet back from a SR-unaware VNF, it has to re-apply the original SR encapsulation (re-attaching the original outer IPv6 header with the SRH) before passing the packet to the IPv6 layer. In other words, the correct VNF chain should be re-associated to the packet. In the general case, a VNF in a NFV node could be inserted in different VNF chains at the same time. Therefore the packets that are forwarded through the VNF should be classified when they go out from the VNF itself to understand to which VNF chain they belong. On the other hand, we can impose the constraint that a VNF in a NFV node can be inserted only in one VNF chain at the same time. Under this constraint, it is possible to operate in a very simple way and associate all packets that go out of the VNF to one VNF chain. To make a concrete example, assume that the packets belonging to a flow $f_1$ are associated by an ingress node to the VNF chain represented by $<$v-a,v-i,v-x$>$, while the packets belonging to the flow $f_2$ are associated to the chain $<$v-b,v-i,v-y$>$. The packets of both flows need to cross the VNF v-i, but those belonging the the flow $f_1$ should be associated to the chain $<$v-a,v-i,v-x$>$ when they go out of the VNF v-i and those belonging to the flow $f_2$ should be associated to the other chain. This scenario is shown in  Fig.~\ref{fig:two-chains}-a. If we duplicate the VNF v-i by instantiating two instances in the same VNF node (v-i1 and v-i2 in Fig.~\ref{fig:two-chains}-b), it is possible to meet the condition that a single VNF is only associated to a VNF chains. The two chains associated to flows $f_1$ and $f_2$ will now be respectively $<$v-a,v-i1,v-x$>$ and $<$v-b,v-i2,v-y$>$.

Formally, let S be the set of all VNF chains allocated in the network. We say that a VNF is \textit{univocally mappable} if it belongs to at most one VNF chain in S. We also assume that a VNF cannot appear twice in the same VNF chain.

For a univocally mappable VNF, it is possible for the SR/VNF connector to associate all the packets outgoing from the VNF to a VNF chain and to the index within the VNF chain. Of course the SR/VNF connector needs to know from which VNF a packet is coming. This knowledge depends on how the internal routing through the VNFs is implemented. For example, if  a virtual switch is used, the VNF identity could be associated to the MAC address of the virtual interface of the VNF. Another example is the solution that we implemented in the NFV node described in the next section, where the VNFs run in separated containers implemented as Linux namespaces. In this case each container has its interface associated to a different virtual interface of the Linux host, therefore the identity of the VNF that processed the packet could be simply obtaining from the virtual interface where the packet comes from.

\begin{figure}[htpb]
	\centering
	\includegraphics[width=0.48\textwidth]{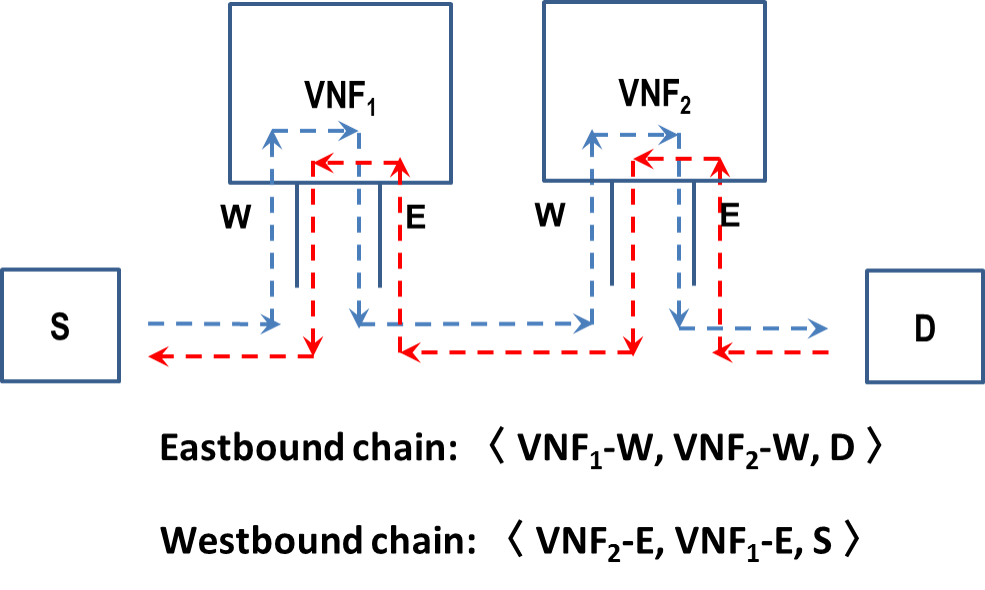}
	\caption{Bi-directional Service Chain example}
	\label{fig:bidir-chain}
\end{figure}

The above described scenario can be extended to the case of bi-directional service chains. A bi-directional services chain is composed of two uni-directional chains, indicated as Eastbound chain and Westbound chain in Fig.~\ref{fig:bidir-chain}. We assume that each VNF which participates in a bi-directional service chain has two interfaces, indicated as W and E in Fig.~\ref{fig:bidir-chain}. Under these assumption, the traffic going out from the E interfaces can be associated to the Eastbound chain (from source node S to dest node D) while the traffic going out from the W interfaces can be associated to the Westbound chain (from D to S). In other words, we assume that each interface of a VNF instance can inject traffic only in one chain at a time. Formally, let S be the set of all VNF chains allocated in the network. Let I be an interface of a VNF instance. We say that an interface I is \textit{univocally mappable} if all the traffic outgoing from I belongs to at most one VNF chain in S. To cover the bi-directiona chains, we now say that a VNF is \textit{univocally mappable} if all its interfaces that output traffic are univocally mappable.

If a VNF is not univocally mappable, a more complex classifier is needed in the SR/VNF connector to associate the packets outgoing from the VNF to the VNF chain and index. We are not covering this general case in the implementation described in the next section and we only provide some architectural considerations hereafter. In general, the classifier could be based on: 
\begin{itemize}
    \item addressing information (IPv6 address, transport protocol and ports);
    \item application level data.
\end{itemize}

Using addressing information could be the simplest way, but it may not work in some application scenarios where different VNF chains are associated to packets belonging to the same transport or application flow (e.g., in case of load balancing).
Mechanisms based on application data may have the same disadvantage and, in addition, they are application-dependent, therefore a specific classifier logic should be deployed in the NFV nodes for each specific application. The SR/VNF connector can also perform operations before forwarding the decapsulated packet to the VNF, like adding specific IPv6 header fields (e.g., flow label or specific IPv6 extension header fields/options) or compute a hash value on the entire packet or on some selected fields, and store the result as a key. It is important to note that these mechanisms may fail when the VNF modifies the packet in the payload and/or header fields, and this has to be taken into account when dealing with SR-unaware network functions.

%% file: inc/implementation.tex
\section{Implementation}
\label{sec:implementation}

In this section a Linux-based implementation of the proposed NFV/SR architecture is presented. In particular, we consider the data plane that is composed of SR edge routers, that classify the traffic and enforce the sequence of  VNFs, and NFV nodes, that are the core routers running the VNFs. Our main focus is the implementation of a NFV node supporting SR-aware and SR-unaware VNFs, with particular emphasis on the mechanisms to support SR-unaware VNFs. In the following, the implementation of the NFV nodes and SR edge routers is separately described. Our implementation is Open Source and available at~\cite{srext-srv6-net-prog}.

\subsection{NFV node}

In order to realize a NFV node, the following main components are required:
\begin{itemize}
    \item An IPv6 layer capable of processing SR-encapsulated packets and passing them to the SR/VNF connector; the actual processing of the the SRH can be performed by the IPv6 layer or by the SR/VNF connector; in the latter case, at least, the IPv6 layer is requested to intercept IPv6 packets and pass them to the SR/VNF connector;
    \item A SR/VNF connector as defined in the previous section, capable of receiving SR packets from the IPv6 layer, processing them, and passing them to the proper VNF;
    \item A dispatching mechanism, able to send packets to the VNFs and to capture the packets returned by the VNFs; the actual way in which the dispatching mechanism is realized is strictly related to how VNFs are implemented;
    \item A virtualization environment, able to run the VNFs as independent processes, or as applications within containers or dedicated Virtual Machines;
    \item The actual VNFs; they can be SR-unaware VNFs, like a standard packet filter, or SR-aware VNFs that are aware of SR and VNF chaining.
\end{itemize}


Starting from the lower layer, we have the IPv6 layer with SR processing. In Linux all networking-related operations, including IPv4 and IPv6 header processing are implemented at kernel-level.
Starting from version 4.10, the Linux kernel supports IPv6/SR (and the corresponding SRH header) based on the implementation provided by the IP Networking Lab at the Universit\'{e} Catholique de Louvain \cite{SRH-linux-impl}. In particular, it is possible to set-up an  IPv6/SR router capable of processing the IPv6 SRHs and forwarding the IPv6 packets with SRH to the next segment.
The Linux kernel implementation can support a scenario in which the VNFs are SR-aware and directly run as IPv6/SR nodes. Instead, in a NFV node in which SR-unaware VNFs run in a virtualization environment on top of a IPv6/SR node, more operations must be executed to support VNF chaining. In particular, IPv6/SR packets should be intercepted and passed to the SR/VNF connector that extracts the original packets and sends them to the proper VNF. 

For these reasons, we designed and implemented some modifications to the default Linux IPv6 packet processing. In Linux, the IPv6 packet handling can be done by directly modifying the IPv6 implementation or by using the \textit{netfilter} framework. Linux netfilter is a modular and powerful framework for packet mangling, offering a number of \textit{hooks} in various points of the Linux kernel network stack that can be exploited to define custom functions.

We preferred to use the latter modular approach and implemented a new kernel module called srext (Segment Routing EXTensions). The srext module is attached to netfilter with a pre-routing hook and acts as SR/VNF connector. All IPv6/SR packet are passed to this SR/VNF connector (at kernel level) for further packet processing according to the presented architecture. Both SR-aware and SR-unaware VNFs are supported. In case of SR-unaware VNFs, the SR encapsulation (and the corresponding SRH) is removed and the original packet is sent to the addressed VNF.

The srext does not only act as SR/VNF connector for the VNF chaining scenario considered in this paper, but it is meant as a full implementation of the SRv6 network programming model~\cite{srv6-net-prog}. The implementation of the srext module is Open Source and available on a public repository~\cite{srext-srv6-net-prog}. 

In our implementation the VNFs run in separated Linux namespaces~\cite{wiki:Linux_namespaces}. The Linux networking stack is used as dispatcher mechanism between VNFs and the SR/VNF connector. Packets are passed to the VNFs and back to the SR/VNF connector through internal virtual interfaces, associated to the different namespaces.
This solution (using the Linux networking as dispatcher) is very simple and is fully compatible with all legacy VNFs that use IP networking as input and output channels. Examples are Firewalls (netfilter/iptables in Linux), Network Address Translators (NATs), or Deep Packet Inspection systems (DPIs). Let us discuss now the cost of this solution of using the Linux forwarding for the exchange of packets from the SR/VNF connector to NFs and vice-versa. We consider a very simple model, and let $f$ be the processing cost of using the networking stack for forwarding a packet. Therefore a normal router has cost $f$, while a NFV node that needs to invoke $n$ SR-aware VNFs for a given packet has cost:

\begin{equation*}
(n+2)f
\end{equation*}

Where $f$ is to forward the packet from SR/VNF connector to the first VNF, $(n-1)f$ is to forward the packet from each VNF to the next one, $f$ is to forward the packet back to the SR/VNF connector, $f$ is to forward the packet to the next hop.

For the case of SR-unaware functions, we need to consider also the processing cost of de-encapsulation and re-encapsulation of the packets. Let us refer to these cost as $d$ and $e$ respectively. The processing cost for a NFV node that needs to invoke $n$ SR-unaware VNFs for a given packet is:

\begin{equation*}
d+(2n+1)f+e
\end{equation*}

Where $d$ is for de-capsulating the packet from the outer IPv6/SRH header, $2f$ is for each two-way exchange between the SR/VNF connector and the $n$ VNFs, $e$ is to re-encapsulate the packet, $f$ is to forward the packet to the next hop.

When only one VNF needs to be invoked for a packet crossing a NFV node, the processing cost in this simple model becomes $3f$ for a SR-aware VNF and $d+3f+e$ for the SR-unaware VNF. The above described model is of course very simplistic. In particular, the main simplification is to assume a fixed cost $f$ for all forwarding operations. In reality, there are several factors that can influence the processing cost of a given packet, belonging to the Link layer or the Network Layer of the protocol stack (for example, the size of the IP forwarding table or the status of the ARP table). 

\subsection{SR edge router}

The objective of an ingress SR edge router in the NFV/SR architecture is to process incoming packets, classify them, and enforce a per-flow VNF chain; the list of VNF identifiers is applied by encapsulating the original packets in a new IPv6 packets with a SRH reporting as segment list the order list of addresses of the given VNFs.

Such SR edge router has been realized using the SRv6 features recently introduced in the Linux kernel (version 4.10) and based on the implementation provided in \cite{SRH-linux-impl}. By using such SRv6 features the ingress SR edge router can classify the traffic and encapsulate the selected packets in a new IPv6 header with the SRH routing extensions. The classification is based on the IPv6 destination addresses. In particular, by using the Linux routing tables in the node it is possible to associate a VNF chain to a set of destination addresses. This is done by creating a IPv6 in IPv6 tunnel in which the matching packets are encapsulated. The SRH routing header is attached to the outer packets of the tunnel and contains the list of SIDs representing the VNFs that should process the packet as well as the egress SR edge router. The egress SR edge routers have to remove the SR encapsulation and forward the inner packet toward its final destination. This allows the final destination to correctly process the original packet. The described classification/encapsulation/decapsulation mechanisms represent the regular behavior of a SR edge router, therefore it was possible to reuse the SRv6 implementation provided in \cite{SRH-linux-impl} with no additional modifications. For simplicity, in our testbed no remote control interface has been added, and the VNF chains are statically configured for each traffic flow using the command line interface (see next section). 

\section{Testing methodology}
\label{sec:methodology}

\subsection{Testbed description and operations}

\begin{figure}[htpb]
	\centering
	\includegraphics[width=0.48\textwidth]{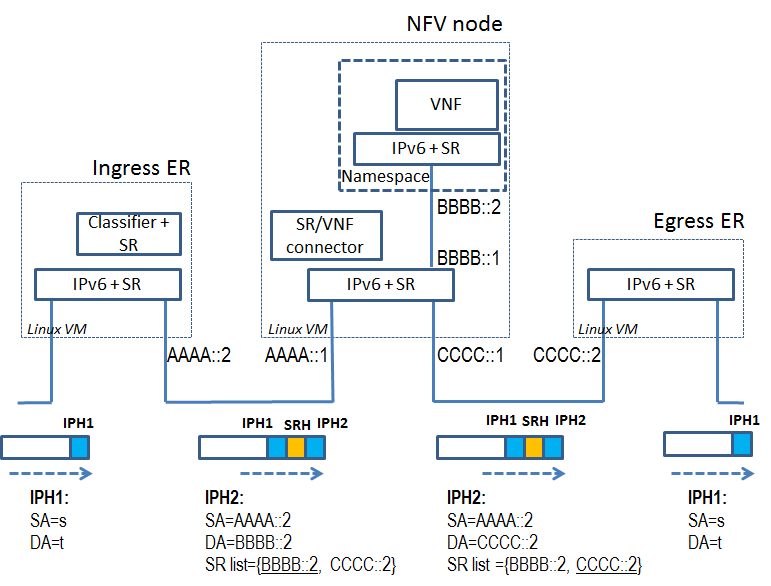}
	\caption{NFV/SR Testbed.}
	\label{fig:testbed}
\end{figure}

In order to verify the correctness of our implementation and to evaluate the performances, we set up simple and easily replicable testbed, shown in Fig.~\ref{fig:testbed}. 

The testbed is composed of three nodes, implemented as Linux Virtual Machines (VMs), running the modified kernel with IPv6/SR support. The advantage of using VMs in place of physical nodes is that it is much simpler to setup (and replicate) the testbed. Moreover, there is the possibility to completely customize the HW and the CPU power of each node.

The testbed is based on the VirtualBox~\cite{virtualbox} virtualization environment and the Vagrant tool~\cite{vagrant} to manage and configure the development of the testbed. The testbed is deployed over a x86 server (AMD Opteron, with 32 cores @2300 MHz). 

The two external nodes in Fig.~\ref{fig:testbed} act as SR edge routers while the central node implements a NFV node. We allocated 8 CPU cores and 16 GB of RAMs to each of the VMs that run the edge routers, and one CPU core and 1 GB of RAM to the VM that run the VNF node. The SR edge routers are implemented as more powerful VMs in order for them to be able to generate a traffic with a rate that can saturate the NFV node (in case of saturation test).

The two ERs (respectively with address \textit{AAAA::2} and \textit{CCCC::2} are connected through two dedicated links to the NFV node (with addresses \textit{AAAA::1} and \textit{CCCC::1}). The network \textit{BBBB::/64} is used for the VNFs running on the NFV node and is present in the routing tables of the two external nodes having as a gateway the NFV node. The iperf traffic sources and sinks have IPv6 addresses \textit{EEEE::2} (source, in the ingress ER) and \textit{DDDD::2} (sink, in the egress ER).

The classifier in the ingress ER is very simple and forces all packets destined to a specific destination address or networks (e.g. \textit{DDDD::2} to traverse a VNF hosted in the NFV node (e.g. \textit{BBBB::2}) before reaching the egress edge node \textit{CCCC::2}. The ER encapsulates packets in new IPv6 packets with SRH with segment list $<$\textit{BBBB::2},\textit{CCCC::2}$>$. A command line example command to configure the classifier and setup the tunnel in the ingress ER is reported hereafter, where \textit{DDDD::2} is the destination prefix, \textit{AAAA::1} the next hop, \textit{encap seg6} tells the Linux kernel to use SR encapsulation instead of normal IPv6, \textit{segs} defines the segment list that will be added in the SRH. 

\texttt{ip -6 route add DDDD::2/64 via AAAA::1 encap seg6 mode encap segs BBBB::2,CCCC::2}  

Within the VM that runs the NFV node, it is possible to deploy a set of VNFs, each one in a different Linux Namespace and with a different IPv6 address (only one VNF is shown in Fig.~\ref{fig:testbed}, with IPv6 address \textit{BBBB::2}).

When a IPv6 packet arrives to our implementation of the NFV node, the packets is processed by our kernel module that is connected to a pre-routing netfilter hook. The module checks if the IPv6 header contains the SRH and the IPv6 destination address corresponds to a SR-unaware VNF hosted in the NFV node (e.g. \textit{BBBB::2}). In this case the encapsulation is removed and the inner packet is internally forwarded to the VNF. In all other cases the packet continues its regular IPv6/SR processing. In particular, if the IPv6 header contains the SRH but the IPv6 destination address corresponds to a SR-aware VNF hosted in the NFV node, the regular IPv6 SR processing will forward the packet to the VNF. 

When a packet comes back from a SR-unaware VNF, it is without encapsulation. The SR/VNF connector re-encapsulate the packet adding the SRH header that represents the VNF chain to which the packet belong. The packet is then passed to the underlying IPv6 layer for being forwarded to the next-hop (the egress edge router in our case). When the egress ER receives the packet, as last segment, it removes the SR encapsulation and forwards the packet toward the destination.

Note that with the same testbed configuration depicted in Fig.~\ref{fig:testbed} we can make experiments with both SR-aware and SR-unaware VNFs, by changing the behavior of the SR/VNF connector in the NFV node and of course the VNF instance that we run in a Namespace in the NFV node.

\subsection{Methodology for performance evaluation}
\label{sec:perfevalmethodology}

In addition to verifying the correct functionality, using the testbed in Fig.~\ref{fig:testbed} we can provide some performance evaluation. Considering that the real performance will depend on the specific hardware configuration, this type of software testbed allows to make a \textit{relative} comparison of different solutions or to study the performance of a given solution with respect to different conditions (e.g. increasing number of flows or number of VNFs to be handled). For example, it is not of interest to evaluate the maximum achievable throughput (packets/s) in absolute terms, but to evaluate the relative difference of this parameter under different conditions.

The methodology is based on traffic generators that can be configured to generate streams of traffic at given packet rates between traffic sources and sinks. Let $R$ be the generated packet rate [packet/s]. Two types measurements are collected: the received packets at the traffic sinks and the CPU utilization in the systems under tests. The received packets are collected by the same tool used to generate the traffic, while a separate tool is needed to keep track of the CPU utilization in the system under test. Let $L(R)$ be the packet loss ratio measured at the traffic sink for a given packet generation rate $R$. We define the packet success ratio $S(R)$ as $1-L(R)$. $U(R)$ is the CPU utilization of the system under test for a packet generation rate $R$. Starting from low packet generation rate $R$, we expect to have a \textit{no loss} region in which $S(R)\approx100\%$ and $U(R)\ll100\%$ (the CPU utilization is well below 100\%). For high packet generation rates we have a \textit{saturation} region in which $S(R)\ll100\%$ (the success ratio is well below 100\%) and $U(R)=100\%$. In between, we have a \textit{transition} region in which the success ratio starts to decrease from 100\% and the CPU utilization starts approaching 100\%. Note that for each rate $R$, we repeat a test of a given duration (e.g. 30 seconds) for a number of times (e.g. 30 runs) in order to evaluate the averages of $S(R)$ and $U(R)$ along with their confidence interval.

Focusing on the \textit{no loss} region, we expect to have a linear increase of the CPU utilization $U(R)$ with respect to input packet rate $R$: $U(R)=mR+k$. By simple linear regression we can estimate the slope $m$ [CPU\%/pps] which represents the CPU load required by the system under test. It is now possible to compare two scenarios or solutions by comparing the evaluated $m$ parameters.

In a virtualized environment like the one used for our experiment, the different nodes and processes may share the same pool of hardware resources. Hence, we need to be careful in order to take significant and accurate measurements of CPU utilization in the part of the system that we want to observe. In particular, the traffic generation tools should be run without interfering with the CPU utilization under measurements (i.e. they should be allocated on different CPU cores). In our experiments described in the next section, we are considering only the performance of the NFV node and are not interested in the ingress and egress Edge Routers. For this reason, we can run the traffic generator tool on the same nodes hosting the ingress and egress ER (respectively, the traffic source in the ingress ER and the traffic sink in the egress ER). 

As packet generation tool we used the \textit{iperf3}~\cite{iperf} application. We found that the number of received packets reported by iperf3 was not accurate (not all received packets were accounted for). Therefore we used the Linux \texttt{ifconfig} command, counting the packets received by the NFV node and transmitted by it. For measuring CPU utilization, we used the Linux \texttt{top} command. Both iperf, ifconfig and top produced log files that we post-processed with our scripts. The top command is not an optimal solution, because we found that it consumes a not negligible amount of CPU, increasing the CPU utilization of the system under test.



%% file: inc/results.tex
\section{Tests and results}
\label{sec:results}

In this section a preliminary performance analysis of our implementation is presented.
The main goal is to estimate the processing overhead introduced by our implementation of a NFV node (supporting SR-unaware VNFs) with respect to a NFV node that just provides routing toward internal VNFs (supporting SR-aware VNFs). We used the testbed described in the previous section and shown in Fig.~\ref{fig:testbed}.

\begin{figure}[htpb]

	\centering
	\includegraphics[width=0.5\textwidth]{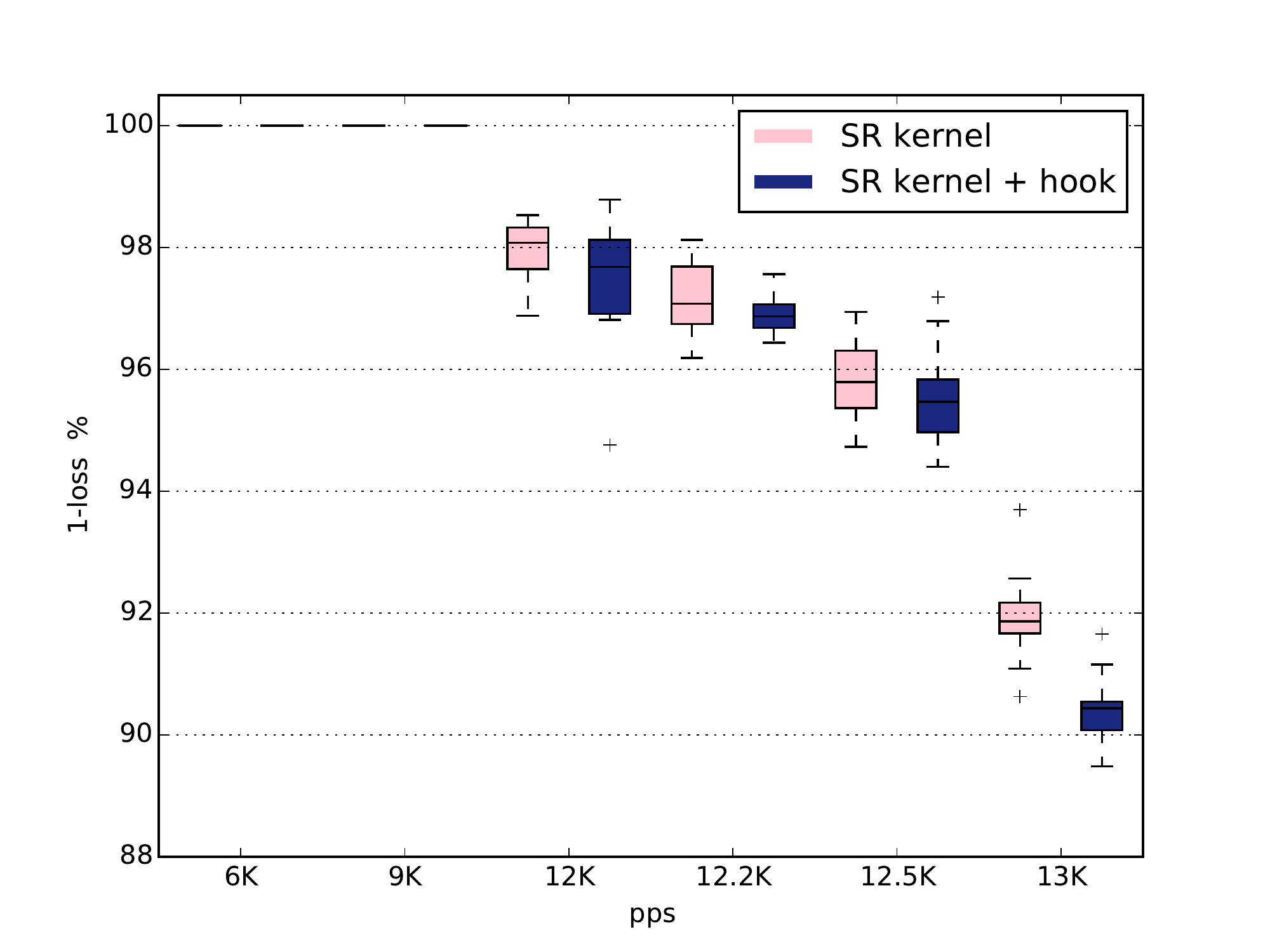}
     \vspace{-2em}

	\caption{Success ratio at different packet rates}
	\label{fig:loss}

\end{figure}

\begin{figure}[htpb]
\vspace{-3em}

	\centering
	\includegraphics[width=0.5\textwidth]{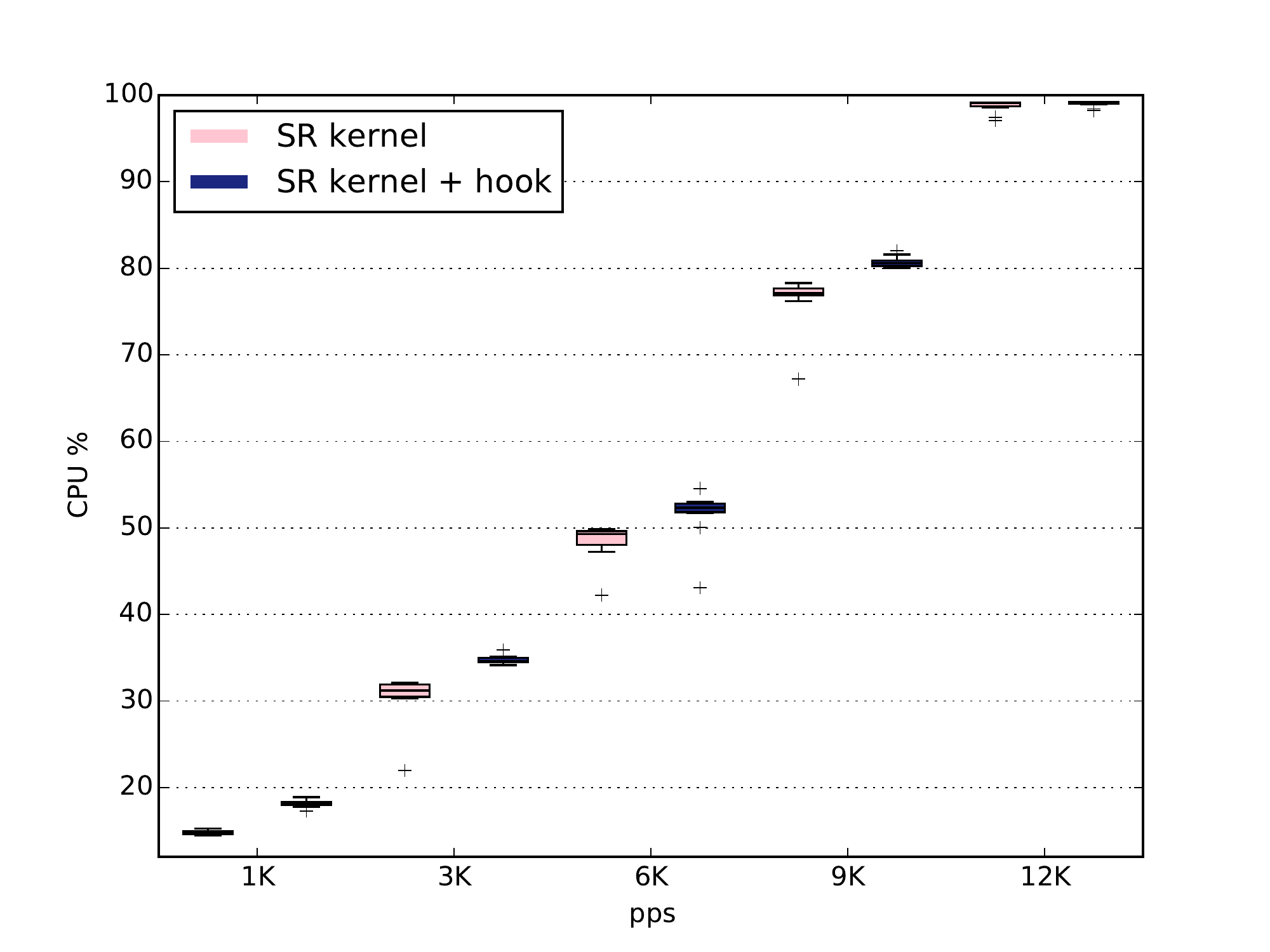}
    \vspace{-2em}

	\caption{CPU Utilization at different packet rates.}

	\label{fig:cpu_2}

\end{figure}

In order to evaluate the performances of the NFV node in comparison with a standard IPv6/SR router, the simplest possible VNF has been considered, i.e. the VNF performs only receives IPv6 packets and performs a routing function, resending back the packets towards the SR/VNF connector. According to the methodology described in the previous section, we generated a flow of UDP packets with payload size equal 1024 bytes sent with rates ranging from 1 to more than 12 kpps. For each packet rate, we consider the basic solution capable of only supporting SR-aware VNFs (denoted as \textit{SR kernel} in the graphs) and our solution to support SR-unaware VNFs (denoted as \textit{SR kernel + hook}). Fig.~\ref{fig:loss} shows the success ratio for different packet rates. For each packet rate, the success ratio of the basic solution is plotted on the left (in red) and our solution on the right (in blue). From 1 kpps to 9 kpps the success ratio is 100\%, therefore we are in the \textit{no loss} region (only 6 kpps and 9kpps are plotted in this region). From 12 kpps onward we have the \textit{saturation} region in which the success ratio is below 100\%. This is confirmed by the CPU utilization results shown in Fig.~\ref{fig:cpu_2}, which show a CPU utilization of around 80\% for the 9 kpps generation rate and of 100\% for 12kpps. Considering the absolute values of achievable throughput in our testbed, we measured a throughput of more than 11kpps in the basic case. We are allocating only a single core to the NFV node and its CPU power is used to perform three routing/forwarding operations per each packet, according to the rough model described in section~\ref{sec:implementation}. Therefore with a single core we could perform a number of forwarding operations per second for IP packets in the order of 33k. For 1500 bytes packets, this would turn out in a throughput of around 400Mb/s.

In the linear region of Fig.~\ref{fig:cpu_2} we can see that the CPU utilization is slightly higher in the \textit{SR kernel + hook} case (corresponding to the support of SR-unaware VNFs) than in the basic case. Likewise, in the saturation region of Fig.~\ref{fig:loss}, we can appreciate a slightly lower success rate for the \textit{SR kernel + hook} case with respect to the basic case.

\begin{figure}[htpb]
\vspace{-3em}
	\centering
	\includegraphics[width=0.5\textwidth]{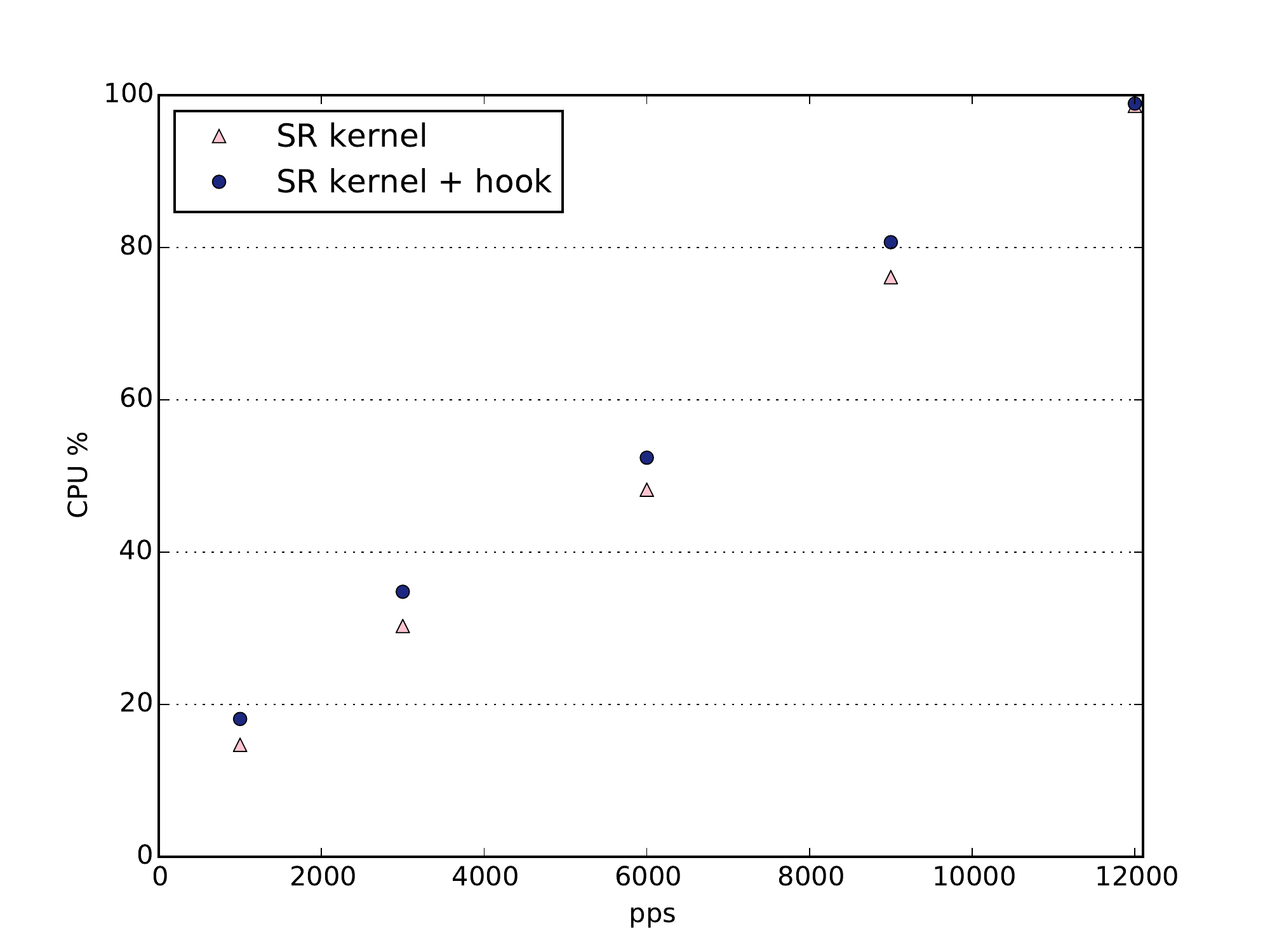}
	\vspace{-2em}

	\caption{CPU Utilization vs packet rate (scatter plot).}

	\label{fig:cpu}
\vspace{-3ex}

\end{figure}

In Fig.~\ref{fig:cpu} we provide a scatter plot of the CPU utilization vs the packet rate, while Table~\ref{tab:tab1} reports the results of the linear regression according to the model presented in sec.~\ref{sec:perfevalmethodology}. We observe that our implementation has an fixed additional CPU overhead ($k$) of 3.6\%, while the additional processing cost depending on the packet rate ($m$) is very limited as it is in the order of 2\%.

\begin{table}[htb]
  \centering
  \begin{tabular}{c | c c}
    \hline
                      & SR & SR + hook \\
    \hline
    k [CPU \%] & 8.9 &  12.5 \\
    m [CPU \%/kpps] & 6.64 & 6.78 \\
    \hline
  \end{tabular}

  \vspace{1em}
  \caption{Estimation of k and m parameters by linear regression}

  \label{tab:tab1}
\end{table}

%% file: inc/soa.tex
\section{Relation with the NSH solution}
\label{sec:soa}

Network Service Header \cite{nsh-id} is used to carry both SFC metadata and path information in an additional header. The traffic steering between VNFs (according to what is specified inside the NSH header) is delegated to other tunneling mechanism like VXLAN, GRE or even SRv6 encapsulation. Traffic is encapsulated in tunnels that goes from one VNF to the next VNF hop, the NSH is added between the tunnel headers and the original packet. The basic mechanism is designed to work with NSH-aware services that are capable to understand and process NSH headers. The largest majority of VNF legacy services is NSH-unaware, in this case before the traffic reaches and after it leaves the VNF a NSH proxy has to process it in order to remove an then reattach the NSH header. Such type of proxy has to remove and reattach the NSH header to each incoming/outgoing packet from the VNF, hence it has to reclassify all the outgoing packets from the VNF.

Note that the considered SRv6 based approach and the NSH approach \cite{nsh-id} are not necessarily mutually exclusive, but there can be coexistence scenarios. In particular, citing \cite{ID-ipv6-spring}, \textit{the use of SRv6 together with the NSH allows building flexible service chains where the topological information related to the path to be followed is carried into the Segment List while the "service  plane related information" (function/action to be performed) is encoded in the metadata, carried into the NSH}. 

NSH is supported in Open vSwitch \cite{ovs} through an unofficial patch \cite{ovs_nsh}. The same patch is used in \cite{odl_sfc} to offer experimental support to NSH in the OpenDaylight \cite{odl} platform. OpenStack Neutron \cite{neutron} is the networking-as-a-service platform used by the OpenStack \cite{ops} project. Currently Neutron does not support service function chaining. There is a proposed SFC API for OpenStack Neutron \cite{neu_sfc}. This API defines a service chain as: i)~Flow classifier - definition of what traffic enters the chain; ii)~An ordered list of Neutron ports that define the chain; iii)~Correlation type - chain metadata encapsulation type. VMs are connected to a Neutron network via Neutron ports, using an ordered chain of ports the traffic is steered through the VM composing the service chain. This makes it possible to create a traffic steering model for service chaining that uses only Neutron ports. This traffic steering model has no notion of the actual services attached to these Neutron ports. The \textit{correlation type} specifies the type of chain correlation mechanism supported by a specific Service Functions (it can be MPLS, NSH, ecc..). This is needed, in case of SFC-aware Service Functions, by the data plane switch to determine how to associate a packet with a chain. In case of SFC-unaware VNFs (VNFs that do not support any correlation mechanism) the correlation type is set to none. The current implementation does not support any correlation mechanism.

%% file: inc/conclusions.tex
\section{Conclusions}
\label{sec:conclusions}

In this paper we have presented a solution for VNF chaining based on IPv6 Segment Routing network programming model and its implementation on a Linux based infrastructure.

In particular, we focused on the support of SR-unaware VNFs, for which the infrastructure needs to remove the Segment Routing encapsulation before forwarding a packet to a VNF and then re-add the encapsulation to continue the processing across the chain. We have provided a Linux kernel module performing the proper adaptation, working for a class of VNFs.

The implementation is based on a Linux kernel module (srext) that supports the IPv6 Segment Routing network programming model, available as Open Source.

We have also provided a methodology for evaluating the performance and used it to evaluate the performance of our module. The preliminary results show that our module is very efficient in the usage of CPU.
